\documentclass[prl,amsmath,twocolumn,showpacs]{revtex4-1}
\usepackage{amsmath,graphicx}
\begin{document}
\def\la{{\langle}}
\def\ra{{\rangle}}
\def\a{{\alpha}}
\def\e{{\epsilon}}
\def\p{\tilde{p}}
\def\x{\tilde{x}}
\def\w{\tilde{W}}
\def\t{\tilde{t}}
\def\k{\tilde{k}}
\def\s{\tilde{\sigma}}
\def\a{\hat{A}}
\def\R{\operatorname{Re}}
\def\I{\operatorname{Im}}
\def\q{\quad}
\def\erfc{\mbox{erfc}}

\title{The Hartman effect and weak measurements "which are not really weak"}
%
%
\author {D.Sokolovski$^{1,2}$ and E.Akhmatskaya$^{3}$}
\address{$^1$ Department of Chemical Physics,  University of the Basque Country, Leioa, Spain\\
$^2$ IKERBASQUE, Basque Foundation for Science\\
$^{3}$ Basque Center for Applied Mathematics (BCAM),\\ Building 500, Bizkaia Technology Park E-48160, Derio, Spain}

   \date{\today}
   \begin{abstract}
 We show that in wavepacket tunnelling  localisation of the transmitted particle amounts to a quantum measurement  of the delay it experiences in the barrier. With no external degree of freedom involved, the envelope of the wavepacket plays the role of the initial pointer state.
Under tunnelling conditions such 'self measurement' is necessarily weak, and the Hartman effect 
just reflects the general tendency of weak values to diverge, as post-selection in the final state becomes improbable. We also demonstrate that it is a good precision, or 'not really weak' quantum measurement:  no matter how wide the barrier $d$, it is possible to transmit
a  wavepacket with a width $\sigma$ small compared to the observed advancement. 
As is the case with all weak measurements, the probability of transmission rapidly decreases with the ratio $\sigma/d$.

\end{abstract}

%
%
\maketitle
\section{I. Introduction}

One often reflects on the controversial nature of the tunnelling time issue.
A common feature of many approaches (for a review see \cite{REVS}-\cite{MUGA}) is that proposed tunnelling times appear as mere parameters,
endowed, unlike most other quantities, with neither probability amplitudes nor
probability distributions.  
Inclusion of such time parameters into the framework of standard quantum theory is clearly desirable.
One such parameter is the phase (Wigner) time used to characterise
transmission of a wavepacket \cite{REVS}-\cite{MUGA}.
In the phase time analysis one typically proceeds by expanding the logarithm of the transmission amplitude $T(p)$ in a Taylor series around the particle's mean momentum $p_0$.
Retaining only linear  terms one finds the transmitted part of a Gaussian pulse to be given by
\begin{eqnarray}\label{z1}
\Psi^T(x,t) \approx
T(p_0) \exp(ip_0\bar{y}) \Psi^0(x-\bar{y},t),\quad
\end{eqnarray}
where $\Psi^0(x,t)$ is the freely propagating 
 state and 
\begin{eqnarray} \label{z2}
\bar{y}(p_0)\equiv i T(p_0)^{-1}\frac{\partial T(p_0)}{\partial p}
\end{eqnarray}
is a complex valued quantity. If spreading of the wavepacket can  be neglected, $\R \bar{y}$
gives the shift of the transmitted pulse $\Psi^T(x,t)$ relative to a freely propagating one, while 
$T(p_0)$ and $\I \bar{y}$ describe overall reduction of its size. The term {\it Hartman effect} (HE)
\cite{HART} refers to the fact that as the width of (e.g., rectangular) barrier $d$ tends to infinity, the tunnelled pulse is advanced roughly by the width of the classically forbidden region $\R \bar{y}\sim d$.
This creates an impression that the barrier has been crossed almost infinitely fast and using the shift to evaluate the duration spent in the region one arrives at the phase time
$\tau_{phase}=(d-\R \bar{y})m/p_0 << md/p_0$, which is independent of the barrier width.
There is a large volume of literature on the HE (see \cite{WIN}-\cite{LUN} and Refs. therein) as well as current interest in its experimental observations \cite{NAT}.
One problem in defining the HE for wavepackets is that one cannot simply fix the shape of the incident pulse and increase the width of the barrier \cite{HART}, \cite{MUGA0}, since eventually the transmission will become dominated by the momenta passing over the barrier, for which Eq. (\ref{z1}) no longer holds.
One can make the pulse ever narrower in the momentum space, but then there is no guarantee that
the spatial width $\sigma$ of the pulse need not become much greater than the advancement one wants to detect.
In this vein, the authors of Ref. \cite{LUN} suggested that the HE is an artifact of the stationary formulation of the scattering theory and cannot be realised once localisation of the tunnelling particle is taken into account. Their conclusions appear to agree with those of Winful \cite{WIN}
who pointed out that in a tunnelling experiment the width of the incident wavepacket must exceed the size of the barrier.
One can, however, imagine an optimal case, where $\sigma$ would be large enough to justify the approximation (\ref{z1}), yet always smaller than the expected advancement $d$. If so, one would be able to observe the advancement associated with the HE 
 in a single tunnelling event.
 \newline
A similar problem arises in the seemingly different context of the so-called weak measurements.
There one measures the value of an operator $\a$ using a pointer whose initial position 
is uncertain, so as not to perturb the measured system. 
If the uncertainty is large, the mean of the meter's readings coincides with the real part of the the weak value of $\a$,
$\la A\ra_W$. This may lie well outside the spectrum of $\a$ or even tend to infinity,
 and one's wish is to observe such an 'unusual' value.
Often the spread of the readings exceeds $Re \la A\ra_W$, thus requiring a large number of trials before the value can be established. If, on the other hand, the large spread can be made significantly smaller than $Re \la A\ra_W$, a single measurement would yield information about $\la A\ra_W$. The authors of \cite{Ah1}, \cite{Ahbook} gave a possible recipe for constructing such  measurements, which they described as 'weak' but not 'really weak'. Analogy between tunnelling times and weak values has been studied in 
 \cite{STEIN}, \cite{AHSOSC} and \cite{S1}-\cite{S2}. Discussion of causality in barrier penetration 
 can be found in \cite{S3}.
 \newline 
 The purpose of this paper is to introduce an amplitude distribution for the phase time $\tau_{phase}$ or, 
 rather for the spacial delay associated with $\tau_{phase}$,
 to demonstrate that locating the transmitted particle amounts to a weak measurement of this delay, 
 and prove that this measurement is of the 'not really weak' kind mentioned above. 
The rest of the paper is organised as follows. In Sections II and III we establish formal equivalence between wavepacket transmission and quantum measurements. In Sect. IV
we use the analogy to analize a weak measurement in the limit where the weak value  tends to infinity. In Sect. V we consider a special case where such a measurement  is not 'really weak'. In Sect. VI
we apply the analysis to the Hartman effect in tunnelling. Sect. VII contains our conclusions.
\section {II. Wavepacket transmission as a quantum measurement.}
Consider a one-dimensional wave packet with a mean momentum $p_0$ incident from the left on a short-range potential $W(x)$.
Its transmitted part is given by (we put $\hbar=1$)
\begin{eqnarray}\label{a1}
\Psi^T(x,t) =\int T(p) C(p)\exp(ipx-i\e(p)t) dp
\end{eqnarray} 
where $p$ is the momentum,  $T(p)$ is the transmission amplitude, $C(p)$ is the momentum distribution of the initial pulse,
and the energy $\e(p)$ is $p^2/2m$ for massive non-realtivistic particles or $cp$ ($c$ is the speed of light) 
for the photons. 
The freely propagating  ($W(x)=0, T(p)=1$) state is given by
\begin{eqnarray}\label{a2}
\Psi^0(x,t) =\int C(p)\exp[ipx-i\epsilon(p)t] dp.
\end{eqnarray}
Writing  $T(p)$
as a Fourier integral,
\begin{eqnarray}\label{a3}
T(p)= \int \xi(y) \exp (-iyp) dp,
\end{eqnarray}
we note that upon transmission an incident plane wave $\exp(ip_0x)$ becomes
a superposition of plane waves with various spacial shifts, $T(p_0)\exp(ip_0x)=
\int dy \xi(y) \exp[ip_0(x-y)]dy$. Thus,
the value of the spacial shift (delay relative to free propagation) with which  a particle with a momentum $p_0$ emerges from a barrier is indeterminate unless the superposition is destroyed.
Such a destruction can be achieved by employing a wavepacket with a finite spacial width.
Inserting Eq. (\ref{a3}) into (\ref{a1}),
we write down the transmitted pulse as a superposition of freely 
propagating states with all possible spacial shifts $y$, 
\begin{eqnarray}\label{a4}
\Psi^T(x,t)=\int \xi(y)\Psi^0(x-y,t) dy.
\end{eqnarray}
If the potential $W(x)$ is a barrier, and, therefore, does not support bound states, 
the causality principle requires
 $\xi(x)$ must vanish for all $y>0$ \cite{S3}.  Thus,
the Fourier spectrum of $T(p)$ (\ref{a3}) contains no positive shifts, 
 and in Eq. (\ref{a4}) 
there are no terms advanced with respect to free propagation.

In the following, we will consider an incident pulse with a Gaussian envelope of a width $\sigma$ and a mean momentum $p_0$, $2/\sigma << p_0$, located at $t=0$  far to the left from the barrier at some $x_0<0$. Thus in Eq. (\ref{a1}) we have
\begin{eqnarray}\label{a5a}
C(p)=\sigma^{1/2}/(2\pi)^{3/4}
\exp [-(p-p_0)^2\sigma^2/4-i(p-p_0)x_0].\quad
\end{eqnarray}
and the free state in Eq. (\ref{a2}) takes the form
\begin{eqnarray}\label{a5}
\Psi^0(x,t)=\exp[ip_0x-i\e(p_0)t] G^0(x,t)
\end{eqnarray}
where $G^0(x,t)$ is a time dependent envelope, whose explicit form is given
in the Appendix A.
 Finally, with the help of (\ref{a5}) and (\ref{a4})  we rewrite Eq. (\ref{a1}) as 
\begin{eqnarray}\label{a6}
\nonumber
\Psi^T(x,t)=\exp[ip_0x-i\e(p_0)t]\int \eta(y,p_0)G^0(x-y,t) dy\\
\eta(y,p_0)=\exp(-ip_0y)\xi(y) \quad .
\end{eqnarray}
Next we demonstrate that finding the transmitted particle at a point $x$ we do, in fact, 
perform a quantum measurement of the shift $y$ for a particle with the momentum $p_0$. In order to do so we compare equation (\ref{a6}) with the one describing a von Neumann quantum measurement on a pre- and post-selected system.

\section {III. Quantum measurement as transmission}

Consider next a freely moving pointer with a position $x$ and an energy $\e(p)=p^2/2m$
The pointer is prepared in a Gaussian state (\ref{a5a}) so that its free evolution is described by Eq. (\ref{a5}). At a time $t=T$ the pointer is briefly coupled to a quantum system which is, at that time, in some state $|\psi_I\ra$.
 Our aim is to measure system's variable represented by an operator $\a$, so that (neglecting for simplicity the system's Hamiltonian) 
we write the total Hamiltonian as
\begin{eqnarray}\label{b1}
\mathcal{H}(t) = -i\partial_x \a\delta(t-t_i) -(2m)^{-1}\partial_x^2
\end{eqnarray}
where $\delta(z)$ is the Dirac delta. 
 After a brief interaction at $t\approx t_i$, the pointer becomes entangled with the system \cite{vN} 
 and at some $t > t_i$ the meter is read, i.e., the pointer position is accurately determined.
 Taking into account the pointer's free evolution, for the state of the composite system $|\Psi(t)\ra$ we have 
\begin{eqnarray}\label{b3}
\la x|\Phi(t)\ra )= \sum_n \la n|\psi_I\ra \Psi^0(x-A_n,T)|n\ra
\end{eqnarray}
where $A_n$ and $|n\ra$ are the eigenvalues and eigenstates of the operator
$\a$, $\a|n\ra=A_n|n\ra$. 
Post-selecting the system in some final state 
 $|\psi_F\ra=\sum_n \la n|\psi_F\ra|n\ra$ 
purifies the state of the meter, which then becomes
\begin{eqnarray}\label{b3}
\Psi^{F\leftarrow I}(x,t)=\exp[ip_0x-i\e(p_0)t]\times
 \\
\nonumber
\int \eta^{F\leftarrow I}(y,p_0)G^0(x-y,t) dy,\quad
\end{eqnarray}
with
\begin{eqnarray}\label{b4}
\eta^{F\leftarrow I}(y,p_0)\equiv\exp(-ip_0y)\times \\
\nonumber
 \sum_n \la \psi_F|n\ra \la n|\psi_I\ra \delta(y-A_n),
\end{eqnarray}
which has the same form as (\ref{a6})
\newline
Defining a state-dependent 'transmission amplitude' 
\begin{eqnarray}\label{b5}
{T^{F\leftarrow I}(p)}\equiv \int \eta^{F\leftarrow I}(y,p_0)\exp[-i(p-p_0)y] dy\\ 
\nonumber
= \sum_n \la \psi_F|n\ra \la n|\psi_I\ra
\exp(-iA_np)
\end{eqnarray}
allows us rewrite Eq. (\ref{b3}) also in a form similar to Eq. (\ref{a1}),
\begin{eqnarray}\label{b6}
\Psi^{F\leftarrow I}(x,t) =\\
\nonumber
\int T^{F\leftarrow I}(p) C(p)\exp[ipx-i\e(p)t] dp
\end{eqnarray} 
where $C(p)$  is given in Eq. (\ref{a5a}). The Fourier series of $T^{F\leftarrow I}(p)$ (\ref{b5}) only contains frequencies $A_n$ from the 
spectrum of $\a$ and, since $T^{F\leftarrow I}(p)$ is a transition amplitude for the Hamiltonian
(\ref{b1}), we have 
\begin{eqnarray}\label{b6a}
| T^{F\leftarrow I}(p)| \le 1.
\end{eqnarray}
Both representations, [(\ref{a1}),\ref{b6})] and [(\ref{a6}),(\ref{b3})], are useful.
Equations (\ref{b3}) and (\ref{a6}) highlight the nature of the measured quantity and the accuracy of the measurement. 
In particular, a von Neumann measurement  the meter determines the value of $\a$ to accuracy $\sigma$. 
If the system is post-selected in $|\psi_F\ra$ and no meter is employed, possible values of $A$, $A_n$, are distributed with probability amplitudes $ \la \psi_F|n\ra \la n|\psi_I\ra$, and the exact value of $A$ remains indeterminate. With the meter switched on, only those values of $A$ which fit under the Gaussian $G^0$ centred  at the observed value $x$  contribute to the amplitude $\Psi^{F\leftarrow I}(x,t)$. Thus, finding the pointer at $x$ guarantees that  $A$ has the value roughly in the interval $[x-\sigma, x+\sigma]$.
Similarly, in Eq. (\ref{a6}) finding the tunnelling particle at a location $x$ determines, to accuracy $\sigma$, 
the delay $y$. 
 Again, for a plane wave with a momentum $p_0$ the value of $y$ is indeterminate,
its amplitude distribution is $\eta(y,p_0)$,
and only those values of $y$ which fit under $G^0$ centred at $x$ contribute to $\Psi^{T}(x,t)$
in Eq. (\ref{a6}). 
\newline
For their part, Eqs.(\ref{a1}) and (\ref{b6}) show that both wavepacket transmission and a quantum measurement
explore local behaviour of the correspondent transmission amplitude, $T(p)$ or $T^{F\leftarrow I}(p)$, in a region of the width $\sigma_p = 2/\sigma$ around $p_0$. They are, therefore, convenient to study the limit in which the momentum width of the initial Gaussian $\sigma_p$ becomes small, i.e., the case of a nearly monochromatic 
initial pulse or an initial meter state broad in coordinate space. 
More discussion of the attributes of the measurement formalism can be found in the Appendix B, 
and in the next Section we consider 
such inaccurate or 'weak' quantum measurements.
\section{IV. A weak quantum measurement}   
If the momentum width of the initial meter's state $\sigma_p$ is small, expanding $\ln T^{F\leftarrow I}(p)$ around $p=p_0$ we arrive at an analogue of Eq. (\ref{z1}),
\begin{eqnarray}\label{c2}
\Psi^{F\leftarrow I}(x,t) \approx\\
\nonumber
T^{F\leftarrow I}(p_0) \exp[i\bar{A}p_0] \Psi^0(x-\bar{A}(p_0),t),
\end{eqnarray}
where 
\begin{eqnarray}\label{c3}
\bar{A}(0)= \bar{A}_1 + i\bar{A}_2\equiv \frac{i}{T^{F\leftarrow I}(p_0)}\frac{\partial T^{F\leftarrow I}(p_0)}{\partial p} =\\
\nonumber
 \int y \eta^{F\leftarrow I}(y,p_0)dy/\int  \eta^{F\leftarrow I}(y,p_0)dy.
\end{eqnarray}
The second equality in (\ref{c3}) defines $\bar{A}(p_0)$ as an 'improper' average \cite{S4} calculated
with the amplitude distribution $\eta^{F\leftarrow I}(y,p_0)$.
For $p_0=0$ the complex valued $\bar{A}(p_0)$  coincides with
 the 'weak value' of $A$, $\la A \ra_W$,  introduced in  \cite{Ah1}  
\begin{eqnarray}\label{c4}
\bar{A}(0)=
  \sum_n A_n \la \psi_F|n\ra \la n|\psi_I\ra/\la \psi_F|\psi_I\ra
=\la A \ra_W.
\end{eqnarray}
Thus, if approximation (\ref{c2}) holds, the final state of the meter is a reduced copy 
of its freely propagating state $\Psi^0$ translated into the complex coordinate plane by $\bar{A}(p_0)$. 
A von Neumann measurement typically employs  a heavy 
pointer at rest, 
prepared in a the Gaussian state  (\ref{a5}) centred at the origin. 
\begin{eqnarray}\label{c4a}
m\rightarrow \infty, \quad p_0=0, \quad x_0=0,
\end{eqnarray}
which will be assumed throughout the rest of this Section.  
With (\ref{c4a}) the Gaussian pointer state (\ref{c2}) becomes
\begin{eqnarray}\label{c5}
\Psi^{F\leftarrow I}(x,t) \approx
\nonumber
K\exp(2i\bar{A}_2x/\sigma^2) \exp[-(x-\bar{A}_1)^2/\sigma^2],\\
K\equiv T^{F\leftarrow I}(0)(2/\pi\sigma^{2})^{1/4}\exp[(\bar{A}_2^2-2i\bar{A}_1\bar{A}_2)/\sigma^2],\quad
\end{eqnarray}
so that the complex translation results in a real coordinate shift $Re \bar{A}$ and a momentum 'kick' of  $2\bar{A}_2/\sigma^2$. This is a known result (see, for example, Ref. \cite{Josz}), and from it we proceed to the main question of this Section.
\newline
It is well known \cite{Ah0}-\cite{Ahbook}that weak values can exhibit 'unusual' properties. For example, $\la A\ra_W$ could be arbitrarily large for initial and final states that are nearly orthogonal,
 $\la \psi_F|\psi_I\ra\approx 0$, even though the spectrum of $\a$ is bounded. We ask  next whether such large shifts can, in principle, be observed with a pointer state whose width $\sigma$ is less than  $Re\la A\ra_W$,
so that the uncertainty in the final pointer position is smaller than the mean measured value?
We note that one can always justify approximation (\ref{c2}) by making the pointer state
narrow in the momentum space, i.e., by sending $\sigma \rightarrow \infty$, but there is no guarantee 
that the spread in the meter reading will not exceed $Re\la A\ra_W$, however large it may be.

As a simple example, consider the case where one measures the $z$-component of a spin $1/2$, $\la A\ra_W=\hat{\sigma}_z,$ pre- and post-selected in the states
$|\psi_I\ra=(|\uparrow\ra+|\downarrow\ra)/2^{1/2}$ and $|\psi_F\ra=[|\uparrow\ra-(1-d^{-1}|\downarrow\ra)]/N^{1/2}$, $N(d)\equiv1+(1-d^{-1})$. Here the parameter $d$ controls the overlap $\la \psi_F|\psi_I\ra$, so that as $d\rightarrow \infty$ 
we have $\la \psi_F|\psi_I\ra\rightarrow 0$. With the help of  Eqs.(\ref{b5}) and (\ref{c3}) we easily find $T^{F\leftarrow I}(p)=[2i\sin(p) -\exp(ip)/d]/(2N)^{1/2}$ and $\bar{A}=2d-1$.
Expanding the logarithm of the transmission amplitude
in a Taylor series around $p=0$,
$ln T^{F\leftarrow I}(p)=\sum_{n=0} (n!)^{-1}\partial ^n \ln T^{F\leftarrow I}(0)/\partial p^n p^n$,
 we note that as $d\rightarrow \infty$,
$\partial ^n \ln T^{F\leftarrow I}(0)/\partial p^n \rightarrow d^n $.  The range of $p$'s contributing to 
the integral (\ref{b6}) is determined by the momentum width of the initial state, $\sigma_p$, so that we have 
$p\lesssim 1/\sigma$. Thus, we can truncate the above Taylor series and, therefore, satisfy the approximation (\ref{c2}), only if $d/\sigma << 1$. 
Consequently, no matter how large the weak value, 
the coordinate  width of the initial pointer state must be even larger.
This weak measurement is, in terms of Ref. \cite{Ahbook}, 'really weak'.

\section{ V. A weak quantum measurement which is `not really weak'}

A 'not really weak' measurement can be realised for a system whose Hilbert space has
sufficiently many dimensions as follows. One can choose initial, $|I\ra$,
  and final, $|F\ra$,  states of the system in such a way that in some vicinity of $p=0$ the transmission amplitude can be approximated as \cite{FOOT}
\begin{eqnarray}\label{c6}
T^{F\leftarrow I}(p)\approx B \exp[-iF(p)d], \quad B=const,\\
\nonumber
F(p)=F_1(p)+iF_2(p), \quad F_2(0) < 0, 
\end{eqnarray}
where $d$ is, as before, a large parameter and $F^{(n)}\equiv \partial^n F/\partial p^n$, $n=0,1,2..$ are all of order of unity. As in the first example, the weak value of $\a$ tends to infinity as $d\rightarrow \infty$,
\begin{eqnarray}\label{c7}
\la A\ra_W= d[F'_1(0)+iF'_2(0)].
\end{eqnarray} 
In Eq. (\ref{b6}) we have $p\lesssim 1/\sigma$, and so may choose 
\begin{eqnarray}\label{c8}
\sigma = \gamma d^{1/2 +\epsilon/2}, \quad \gamma < 1=const(d),\quad 0<\epsilon\le 1,
\end{eqnarray}
so that the the width $\sigma$, although large for a large $d$, is always smaller than 
$Re \la A\ra_W=F'_1(0)d$. (For $\epsilon=1$
we have $\sigma/Re \bar{A} = \gamma$, otherwise $\lim_{d\rightarrow \infty} \sigma/Re \bar{A}=0$.) Returning to 
the Taylor expansion of $\ln T^{F\leftarrow I}(p)$, $-iF(p)d=-id\sum_{n=0}(n!)^{-1} F^{(n)}(0)p^n$
we note that whilst the first two terms are proportional to $d$ and $d^{1/2-\epsilon/2}$ respectively,
the higher order terms behave as $d^{1-n(\epsilon+1)/2}$, $n=2,3,..$, and can, therefore, be neglected as $d\rightarrow \infty$.  With this the Gaussian meter state (\ref{c5}) becomes
\begin{eqnarray}\label{c9}
\nonumber
lim_{d\rightarrow \infty}\Psi^{F\leftarrow I}(x,t) \approx
K \exp[2ixF'_2(0)/(\gamma^2d^{\epsilon})]\\
\nonumber
\exp\{-[x-dF'_1(0)]^2/\gamma^2d^{1+\epsilon}\},\\
 K=B [2/(\pi \gamma^2d^{1+\epsilon})]^{1/4}\times \quad\quad\quad\quad\quad\quad\quad\quad\\
 \nonumber
 \exp\{-iF(0)d+d^{1-\epsilon}[2iF'_1(0)F'_2(0)+\bar F'_2(0)^2]/\gamma^2]. 
\end{eqnarray}
In Eq. (\ref{c9}) we have, apart from a constant and a phase factor, a reduced copy of the original Gaussian shifted 
 by a distance exceeding its width. This weak measurement is, therefore, 'not really weak'.
 
 Finally, to demonstrate that as $d\rightarrow \infty$, $\Psi^{F\leftarrow I}$
 does indeed build up from the momenta 
 in an ever narrower vicinity of $p=0$, it is helpful to evaluate the contribution to the integral (\ref{b6})
 from the tail of the momentum distribution $A(p)$, i.e., from $p$'s greater then some fixed
 $p_{min}$, 
\begin{eqnarray}\label{c10}
I\equiv |\int_{p_{min}}^{\infty}\exp(-p^2\sigma^2/4) T^{F\leftarrow I}(p)\exp(ipx)dp|\quad\\
\nonumber \le \int_{p_{min}}^{\infty}\exp(-p^2\sigma^2/4) |T^{F\leftarrow I}(p)|dp \\
\nonumber
\le
\int_{p_{min}}^{\infty}\exp(-p^2\sigma^2/4)dp = \pi^{1/2}\sigma^{-1} \erfc(\sigma p_{min}/2),
\end{eqnarray} 
 where $\erfc (z)$ is the complementary error function, and we have used Eq. (\ref{b6a})
in going from the second inequality to the third. Using the large argument asymptotic 
of $erfc (z)$ shows that as $d\rightarrow \infty$,
\begin{eqnarray}\label{c11}
I\sim (\gamma^2 p_{min}d^{1+\epsilon})^{-1}\exp(-d^{1+\epsilon}\gamma^2 p_{min}^2/4).
\end{eqnarray}
For any $0<\epsilon\le 1$, and $d\rightarrow \infty$,  $I$ can, therefore, be neglected in comparison 
with $\Psi^{F\leftarrow I}(x)\sim d^{-(1+\epsilon)/4}\exp[-|F_2(0)|d]$, which proves the above point.

Rather than proceed with the construction of a weak von Neumann measurement corresponding to (\ref{c6}), we continue with the analysis of tunnelling across a potential barrier, equivalence between the two cases  demonstrated in Sections II and III.

\section{IV Hartman effect with wavepackets}
Consider tunnelling of a Gaussian wavepacket  (\ref{a5}) representing non-relativistic particle of unit mass, $m=1$, across a rectangular potential barrier of a height $W$ and width $d$. The frequently quoted transmission amplitude is given by 
 \begin{equation}\label{d1}
T(p)=\frac{4ip\kappa \exp(-ipd)}{(p+i\kappa)^2\exp(\kappa d) -(p-i\kappa)^2\exp(-\kappa d)}, 
\end{equation} 
where $\kappa=(2W-p^2)^{1/2}$. For a fixed initial momentum $p_0$ and a height $W$, $p_0^2/2m < W$, we increase the barrier width $d$ in order observe the advancement of the transmitted pulse relative to free propagation.
As in the case of a `not really weak' quantum measurement we wish to find a condition
where the advancement exceeds the wavepacket width by as much as possible. 
As $d\rightarrow \infty$, and for $p\approx p_0$ we have 
\begin{eqnarray}\label{d2}
T(p) \approx  
 \frac{4ip_0\kappa_0 }{(p_0+i\kappa_0)^2}\exp[-i(p-i\kappa)d],\\
 \nonumber
  \kappa_0= (2W-p_0^2)^{1/2},\quad
 \end{eqnarray}
which clearly has the form (\ref{c6}) with $F(p)=[p-i\kappa(p)]d$.
 and we can use Eq. (\ref{c8}) to define the wavepacket 
width $\sigma$ in such a way that it increases with the barrier width $d$ while always remaining smaller than $d$. The only difference with the case analysed in Sect.V is that now the tunnelling particle plays the role of a pointer whose mass, $m=1$, and the mean momentum $p_0$  are both finite. In particular, in the language of measurement theory,
Eq. (\ref{z2}),
\begin{eqnarray}\label{d2a}
\bar{y}(p_0)\sim d+ip_0d/\kappa_0,\quad 
 \end{eqnarray} gives the weak value of the shift relative to free propagation, experienced by a tunnelling particle
 with momentum $p_0$.
 The weak value is of an `unusual' kind mentioned above: $Re\bar{y}(p_0)\sim d$ lies far beyond the range 
 $-\infty <y\le 0$ allowed by causality \cite{S3}.
 
We note further that in Eq. (\ref{c2}) spreading of the wavepacket $\Psi^0(x,t)$ can be neglected, 
since spreading results in replacing initial width $\sigma$ by a complex time dependent width 
$\sigma_t^2 \equiv (\sigma^2+2it)$ [cf. Eq. (\ref{P1})].
We wish to compare positions of the freely propagating and tunnelling pulse roughly at the time it 
takes the free particle to cross the barrier region, i.e., at $t \sim d/p_0$. Thus, as $d\rightarrow \infty$ we have 
$\sigma_t^2\approx \gamma^2 d^{1+\epsilon}[1+2i/(d^{\epsilon}p_0)] \sim \gamma^2 d^{1+\epsilon}=\sigma^2$.
Equation (\ref{c2}) for  the transmitted  Gaussian wavepacket now reads 
 \begin{eqnarray}\label{d3}
\Psi^T(x,t)\approx K\exp[i(p_0x-i\epsilon(p_0)t]\exp[2ip_0X/(\kappa_0 d^{\epsilon}\gamma^2)]\quad\quad\\
\nonumber
 \exp(-(X-d)^2/(\gamma^2d^{1+\epsilon})],\quad\quad\\
\nonumber
K=T(p_0)[2/(\pi \gamma^2d^{1+\epsilon})]^{1/4}\exp[(p_0^2/\kappa_0^2-2ip_0/\kappa_0)d^{1-\epsilon}/\gamma^2]
\end{eqnarray} 
where
\begin{eqnarray}\label{d3a}
X(x,p_0,x_0,t)\equiv x-p_0t-x_0
 \end{eqnarray} 
 is the particle's position relative to the centre of the freely propagating pulse.
 It is readily seen that the peak of the  transmitted density, $\rho^T(x,t) \equiv |\psi^T(x,t)|^2 \approx
 |K|^2 \exp(-2(X-d)^2/(\gamma^2d^{1+\epsilon})]$ is advanced, as desired,
 by a distance $\sim d$ exceeding its width $ \sim \gamma d^{(1+\epsilon)/2}/\sqrt{2}$.
Note that Eqs.(\ref{c10}) and (\ref{c11}) guarantee that the contributions from the momenta 
passing above the barrier, $p>p_{min}=\sqrt{2W}$, are negligible and the transmission 
is always dominated by tunnelling, rather than by the momenta passing above the barrier.
\newline
The advancement mechanism relies on the wavepacket exploring local behaviour of the transmission amplitude $T(p)$ which oscillates around $p= p_0$ with the period
 $\tau_p=2\pi/Re \bar{y}(p_0) = 2\pi/d$. The number of times $T(p)$ must oscillate 
within the momentum width of the pulse $\sigma_p=2/\sigma$ in order 
 to ensure a given ratio of the spacial width of the pulse to the observed advancement,
 $\nu = \sigma/d$, is given by 
\begin{eqnarray}\label{d4}
n_{osc} \equiv \sigma_p/\tau_p= 1/\pi\nu=d^{(1-\epsilon)/2}/\pi \gamma.
 \end{eqnarray} 
The cost of advancement  (\ref{d3}) in terms of the tunnelling probability $P^T$ for a particle
with an energy equal to half of the barrier height, 
$P^T \equiv \int |\psi^T (x,t)|^2 dx \sim |T(p_0)|^2\sim \exp(-2W^{1/2}d),$
 can be estimated by
recalling that the approximation (\ref{z1}) requires $|2d\sigma^{-2}F''(p_0)|=2d\sigma^{-2}\partial^2 \kappa(p_0)/\partial p^2 = 4(d/\sigma)^2/(W^{1/2}d)<<1$. Thus, for a bound on $P^T$, given 
the value of the ratio $\nu$, we have
\begin{eqnarray}\label{d5}
P^T(\nu) << \exp(-8/\nu^2).
\end{eqnarray} 
\newline
We conclude by giving in Fig.1 a comparison between the exact  (\ref{a1}) and approximate  (\ref{d3}) forms of the tunnelled wavefunction for a broad, $W^{1/2}d>>1$, rectangular potential barrier.
\begin{figure}[h]
\includegraphics[width=9cm, angle=-0]{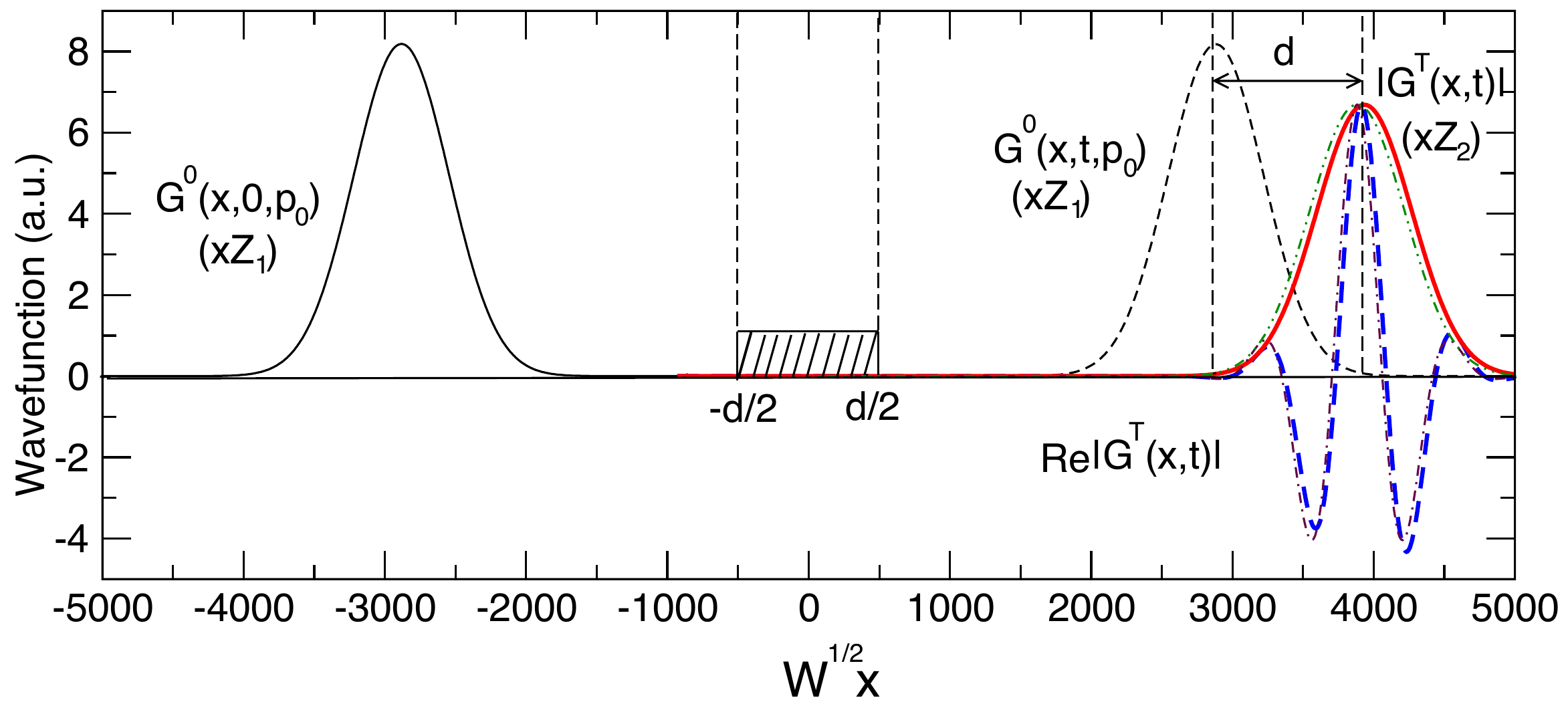}
\vskip0.5cm
\caption{Modulus (thick solid) and real part (thick dashed) of the transmitted pulse, $G^{T}(x,t)\equiv
\exp[-ip_0x-i\epsilon(p_0)t]\psi^T(x,t)$, multiplied by a (large) factor  $Z_2=exp[i(p_0-ik_0)d]$.
The same quantities evaluated using Eqs.(\ref{d3}) are shown by the dot-dot- and dot-dashed lines, respectively. Also shown are the initial (solid) and final (dashed) free envelopes (\ref{P1})
multiplied by a factor $Z_1=200.]$. Other parameters are $W^{1/2}d=10^3$, $\epsilon=0.85$,
$W^{1/2-\epsilon}\gamma=0.8$, $W^{-1/2}p_0=1$,$Wt= W(d+10\sigma)/p_0=5765.$
}
\label{fig:FIG1}
\end{figure}

\section{VII. Conclusions}
In summary, spacial delay in transmission is conveniently analysed in terms of quantum measurement theory.
Classically, a particle with a momentum $p_0$ passing over a potential barrier experiences a unique delay $y$ (an advancement if $y>0$) relative to the free propagation.
This delay can, if one wishes, be used to determine the duration $\tau$ the particle has spent in the barrier region.
Quantally, there is no unique spacial delay for tunnelling with a momentum $p_0$, but rather a continuous spectrum of possible delays, which for a barrier extends from $-\infty$ to $0$,
$-\infty <y\le 0$.
Finding the particle with a mean momentum $p_0$ at a location $x$ one effectively performs  a quantum measurement of the delay $y$.
This is evident from comparing the state for the transmitted pulse (\ref{a6}) and the final state 
of the pointer in a von Neumann measurement with post-selection (\ref{b3}).
The physical conditions are clearly different.
A von Neumann measurement requires coupling to an additional (pointer) degree of freedom, 
while a tunnelling particle 'measures itself', with the role of the initial meter state played by the 
envelope of the wavepacket, superimposed on a plane wave with momentum $p_0$.
Yet a further analogy is valid. 
Just as the width of the initial pointer state determines the accuracy of a von Neumann measurement, the spacial width of the incident wavepacket $\sigma$
determines the uncertainty with which one can know the delay.
 For a large $\sigma$, the measurement is weak and is, therefore, capable of producing 'unusual' results outside the spectrum of available delays.
Spacial advancement of the transmitted pulse, 
while causality limits the spectrum of delays to $y\le 0$,  is just another example of such an 'unusual' value. Quantally, there is no reason for converting it into an estimate for the duration spent in the barrier or sub-barrier velocity. Yet even if such a conversion is made, a tunnelling 
electron can be said to 'travel as a speed greater than $c$' no more than a spin $1/2$ for which 
a weak measurement of the $z$-component finds value of $100$ \cite{Ah0} can be said 'to be a spin $100$'.

Mathematically, the Hartman effect reflects a general property of the weak values which become infinite 
as the probability to reach the final (in this case, transmitted) state vanishes (cf. Eq. (\ref{c4}). In Section VI
we have demonstrated that the measurement of the delay as the barrier width $d\rightarrow \infty$
is not just weak, but also belongs to the class of good precision measurements 
which `are not really weak'. Contrary to the suggestions of Refs. \cite{WIN}, \cite{LUN}, if the barrier is broad, one can always find a wavepacket with a width large but smaller that $d$, which would 
tunnel and exhibit an advancement  by approximately the barrier width.
This is a consequence of the exponential behaviour of the transmission amplitude in Eq. (\ref{d2})
and some properties specific to a Gaussian wavepacket.
We note also, that in our estimate the width $\sigma$ must increase with $d$ at a rate no slower than $\sim d^{1/2}$.
This agrees with the findings of Refs. \cite{MUGA0}, whose authors analysed the Hartman effect using flux-based arrival times.
It also agrees qualitatively with the best relative uncertainty of above $1/N^{1/2}$, achievable in a weak measurement on a system consisting of a large number, $N>>1$, of spins $1/2$ \cite{Ah1}, \cite{Ahbook}. Thus, just as in the case of a good precision weak measurement, a single tunneling event may suffice to observe the Hartman effect. One would also have to wait a long time for that single event, as the tunnelling probability rapidly decreases with the decrease of the ratio $\sigma/d$. 
We will follow the authors  of \cite{Ahbook} in assuming that whoever might perform the experiment is sufficiently patient and has time on his/her hands.

\section{Acknowledgements}
One of us (DS) acknowledges support of  the Basque Goverment grant IT472
and MICINN (Ministerio de Ciencia e Innovacion) grant FIS2009-12773-C02-01.

\section{Appendix A. Some Gaussian integrals.}
Evaluating Gaussian integral (\ref{a2}) with $C(p)$ given by Eq. (\ref{a5}) yields
\begin{eqnarray}\label{P1}
G^0(x,t,p_0)=[2\sigma^2/\pi\sigma_t^4]^{1/4}\exp[-(x-p_0t - x_0)^2/\sigma_t^2]\quad\quad
\end{eqnarray}
where $\sigma_t^2 \equiv (\sigma^2+2it)$ for a particle of a unit mass, $\epsilon(p)=p^2/2$.
For a photon, $\epsilon(p)=cp$, we have
\begin{eqnarray}\label{P1}
G^0(x,t,p_0)=[2/\pi\sigma^2]^{1/4}\exp[-(x-ct - x_0)^2/\sigma^2].\quad\quad
\end{eqnarray}

\section{Appendix B. Connection with the two-state formalism.}
In \cite{Ahbook} (see also Refs. therein)  the authors have formulated a time-symmetric description for a system pre- and post-selected in the states $|\psi_I\ra$ and  $|\psi_F\ra$ at some times $t_1$ and $t_2$, respectively. Evolving $|\psi_I\ra$ and  $|\psi_F\ra$ forwards and backwards in time to the moment
$t_i$ when the system interacts with an external meter, yields what the authors of \cite{Ahbook} called a two-state vector,
$\la\hat{U}^{-1}(t_2,t_i)\psi_F| |\hat{U}(t_i,t_1)\psi_I\ra$, where $\hat{U}(t,t')$ is the system's evolution operator. The vector contains sufficient information to describe the statistical properties of the observed system at $t_i$ \cite{Ahbook}. For example, for $\hat{U}\equiv 1$
the weak value (\ref{c4}) of an operator $\a$ takes the simple form
 $A_W=\la\psi_F|\a|\psi_I\ra/\la\psi_F|\psi_I\ra$. 
 A similar description can also be applied to the case of barrier penetration.  One recalls that $T(p_0)$ is a transmission amplitiude for a particle pre-selected in the plane wave $|p_0\ra$ travelling to the right, $\la x|p_0\ra=\exp(ip_0x)$, at some $t_1$ in the distant past, and then post-selected in the same state at some $t_2$ in the distant future, 
\begin{eqnarray}\label{B1}
\exp[-i\epsilon(p_0)(t_2-t_1)] T(p_0)=\la p_0|\hat{U}(t_2,t_1)|p_0\ra,
\end{eqnarray} 
where $\hat{U}(t_2,t_1)$ may include effects of adiabatic switching of the barrier potential. The post-selection excludes the possibility of reflection, i.e. the particle ending up in the state $|-p_0\ra$. With $|\psi_I\ra$ and  $|\psi_F\ra$ thus defined, one can introduce a two-state vector $\la\hat{U}^{-1}(t_2,t)p_0| |\hat{U}(t,t_1)p_0\ra$ for any $t_1\le t\le t_2$. However, immediate advantage of such a description is not clear since transmission, unlike an impulsive von Neumann interaction,  is a continuous process and no simple expression, e.g., for the weak shift (\ref{z2}), is obtained as a result.



\begin{thebibliography}{999}
\bibitem{REVS} For reviews see: E.H. Hauge and J.A. Stoevneng,
Rev. Mod. Phys. \textbf{61}, 917 (1989); C. A. A. de Carvalho, H. M. Nussenzweig,
Rev. Mod. Phys. \textbf{364}, 83 (2002); V. S. Olkhovsky, E. Recami and J. Jakiel,
Rev. Mod. Phys. \textbf{398}, 133 (2004)
\bibitem{WIN} H. G. Winful, Phys. Rep.\textbf{436}, 1 (2006)
\bibitem{MUGA} J. G. Muga in {\it Time in Quantum Mechanics.  Vol.1}, Second Edition, ed. by. G. Muga,
R. Sala Mayato and I. Egusquiza, (Springer, Berlin Heidelberg, 2008)
\bibitem{HART} T. E. Hartman, J. Appl. Phys., {\bf 33}, 3427 (1962)
\bibitem{MUGA0} S. Brouard. R. Sala, J. G. Muga and I. L. Equsquiza,
Phys. Rev.A, \textbf{49}, 4312 (1994)
\bibitem{MUGA1}V. Delgado and  J. G. Muga, Ann. Phys. \textbf{248}, 122 (1996)
\bibitem{MUGA2} J. G. Muga, I. L. Equsquiza, J. A. Damborenea and F. Delgado,
Phys. Rev. A, \textbf{66}, 042115 (2002)
\bibitem{LUN} J. T. Lunardi, L. A. Manzoni and A. T. Nystrom,
Phys. Lett. A, \textbf{375}, 415 (2011)
\bibitem{NAT} M. D. Stenner, D. J. Gauthier, and M. A. Neifeld, Nature (London) \textbf{425}, 695 (2003)
\bibitem{Ah0} Y. Aharonov, D. Albert and L. Vaidman, Phys.Rev.Lett, {\bf 60}, 1351 (1988)
\bibitem{Ah1} Y. Aharonov, J. Anandan, S. Popescu and L. Vaidman, Phys.Rev.Lett, {\bf 64}, 2965 (1990)
\bibitem{Ahbook}Y. Aharonov and L. Vaidman in {\it Time in Quantum Mechanics.  Vol.1}, Second Edition, ed. by. G. Muga,
R. Sala Mayato and I. Egusquiza, (Springer, Berlin Heidelberg, 2008)
\bibitem{Josz} R. Jozsa, Phys. Rev. A, {\bf 76}, 044103 (2007)
\bibitem{INT}P. B. Dixon, D. J. Starling, A. N. Jordan and J. C. Howell Phys. Rev. Lett.  \textbf{102}, 173601 (2009)
\bibitem{POP} S. Popescu, Physics \textbf{2}, 32, (2009)
\bibitem{STEIN}A. M. Steinberg, Phys. Rev. Lett. \textbf{74}, 2405 (1995)
\bibitem{AHSOSC}  Y. Aharonov, N. Erez and B. Reznik,
Phys. Rev. A, {\bf 65}, 052124 (2002); J. Mod. Opt. \textbf{50}, 1139 (2003)
\bibitem{S1} D. Sokolovski, A. Z. Msezane and V. R. Shaginyan,
Phys. Rev. A, {\bf 71}, 064103 (2005)
\bibitem{S2} D. Sokolovski and R. Sala Mayato,
Phys. Rev. A, {\bf 81}, 022105 (2010)
\bibitem{S3} D. Sokolovski,
Phys. Rev. A, {\bf 81}, 042115 (2010)
\bibitem{vN}J. von Neumann, {\it Mathematical Foundations of Quantum Mechanics } (Princeton University Press, Princeton, NJ, 1955)
\bibitem{S4} D. Sokolovski,
Phys. Rev. A, {\bf 76}, 042125 (2007)
\bibitem{FOOT} This can be done \cite{Ah1},\cite{S2}, for example, by 
choosing the states 
$|\psi_I\ra$ and $|\psi_F\ra$ in such a way that $K$ first moments of $\eta^{F\rightarrow I}$
in Eq. (\ref{b4}) coincide with $T^{(n)}(p_0)/T(p_0)$, $T(p)$ given by Eq,(\ref{c6}). 
\end{thebibliography}
\end{document}